\documentclass[aps,pre, preprintnumbers ,floats,twocolumn,superscriptaddress]{revtex4}
\usepackage{graphicx}% Include figure files
\usepackage{dcolumn}% Align table columns on decimal point
\usepackage{bm}% bold math
\usepackage[english]{babel}
\usepackage[latin1]{inputenc}
\usepackage{graphicx}
\usepackage{epstopdf}
\usepackage{amsfonts}
\usepackage{color}
\usepackage{epsfig}
\usepackage{mathrsfs}
\usepackage{amsmath,amssymb}
\usepackage{subfigure}

\usepackage{framed}
\usepackage[svgnames]{xcolor}
\renewcommand{\(}{\left(}
\renewcommand{\)}{\right)}

\newcommand{\laplinter}{L^{I}}

\newcommand{\laplintra}[1]{L^{(#1)}}
\newcommand{\laplmultinter}{\mathcal{L}^{I}}
\newcommand{\laplmultintra}{\mathcal{L}^{L}}

\newcommand{\supralapl}{\mathcal{L}}

\newcommand{\ds}{\displaystyle}
\newcommand{\module}[1]{|| {#1} ||}

\date{}

\begin{document}

%\graphicspath{{figure/}}
\selectlanguage{english}

%\preprint{\emph{URV pre-print}}

\title{Spectral properties of the Laplacian of multiplex networks}

\author{A. Sol\'e-Ribalta}
  %%\homepage{}
  \affiliation{Departament d'Enginyeria Inform\'atica i Matem\'atiques, Universitat Rovira i Virgili, Tarragona, Spain}

\author{M. De Domenico}
  %%\homepage{}
  \affiliation{Departament d'Enginyeria Inform\'atica i Matem\'atiques, Universitat Rovira i Virgili, Tarragona, Spain}

\author{N. E. Kouvaris}
\affiliation{Departament de F\'{\i}sica Fonamental, Universitat de Barcelona, Barcelona, Spain}

\author{A.  D\'{\i}az-Guilera}
\affiliation{Departament de F\'{\i}sica Fonamental, Universitat de Barcelona, Barcelona, Spain}

\author{S. G\'omez}
  \affiliation{Departament d'Enginyeria Inform\'atica i Matem\'atiques, Universitat Rovira i Virgili, Tarragona, Spain}

\author{A. Arenas}
  \affiliation{Departament d'Enginyeria Inform\'atica i Matem\'atiques, Universitat Rovira i Virgili, Tarragona, Spain}

\date{\today}

\begin{abstract}
One of the more challenging tasks in the understanding of dynamical properties of models on top of complex networks is to capture the precise role of multiplex topologies. In a recent paper, G\'omez~et~al.\ [Phys. Rev. Lett. {\bf101}, 028701 (2013)] proposed a framework for the study of diffusion processes in such networks. Here, we extend the previous framework to deal with general configurations in several layers of networks, and analyze the behavior of the spectrum of the Laplacian of the full multiplex. We derive an interesting decoupling of the problem that allow us to unravel the role played by the interconnections of the multiplex in the dynamical processes on top of them. Capitalizing on this decoupling we perform an asymptotic analysis that allow us to derive analytical expressions for the full spectrum of eigenvalues. This spectrum is used to gain insight into physical phenomena on top of multiplex, specifically, diffusion processes and synchronizability.
\end{abstract}

\maketitle

\flushbottom

%%%%%%%%%%%%%%%%%%%%%%%%%%%%%%%%%%%%%%%%%%%%%%%%%%%%%%%%%%%%%%%%
\section{Introduction}

\begin{figure}[tbp]
  \begin{center}
  \mbox{\includegraphics[width=\columnwidth]{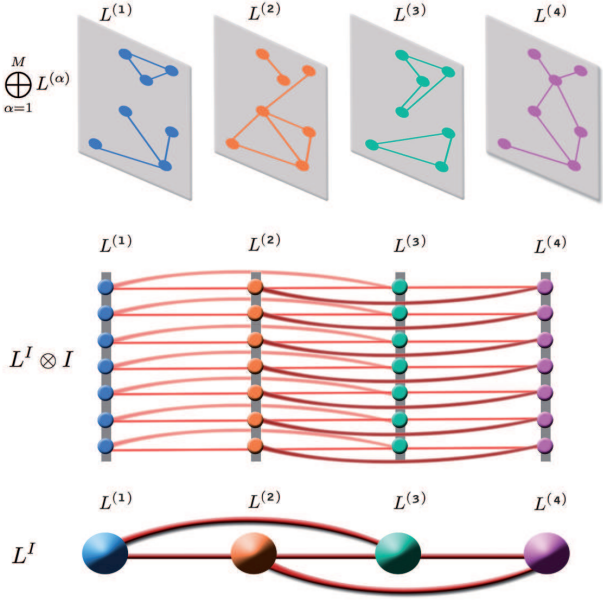}}
  \end{center}
  \caption{Sketch of a multiplex networks of four layers. Note that the same nodes are represented at each layer, although with different connectivity in each one of them. The interlayer connections corresponds to the links between the different categorical layers each node has. In the upper panel we observe the elements that will end in the direct sum in Eq.\ref{eq:intraLaplacian}. In the middle panel we show the structure corresponding to Eq.\ref{eq:interLaplacian}. Finally, in the lower panel, we show the structure that will correspond to the network of layers and its laplacian $\laplinter$.}

  \label{fig:multiplex}
\end{figure}

One of the major lines of research in complex networks has focused in the comprehension of the relationship between network topologies and the behavior of processes occurring on them. The general approach to model a network is not general enough to ascertain the true interdependence that arises in some complex systems. Examples of such a systems are: user relationships in different social networks \cite{magnani11}, transportation systems \cite{cardillo13}, or the learning organization in the brain \cite{basset10}. Particularly interesting are those topologies of interconnected networks called multiplex, see Fig.~\ref{fig:multiplex}, where each object is univocally represented in each independent layer and so the interconnectivity pattern among layers become one-to-one \cite{kurant06,mucha10,Szell10,gardenes12,baxter12,cozzo12,bianconi13}, allowing the simultaneous study of different interconnected patterns between the same objects.

The behavior of any linearized dynamical process on a complex system is related to the Laplacian matrix of the underlying network and particularly to its second smallest eigenvalue. This eigenvalue, also called algebraic connectivity $\lambda_2$, turns out to be essential to understand, for example, the time required to synchronize phase oscillators \cite{almendral07}, or to converge to the maximum entropy state in a diffusion process \cite{gomez13}. Moreover, the largest eigenvalue of the Laplacian matrix plays a determinant role in the assessment of the stability of the synchronization manifold in networks of coupled oscillators \cite{pecora98,arenas08}. In this article, we rely on the particularities of the multiplex networks to model its Laplacian matrix in terms of a decomposition between intra- and interlayer structure.  This decomposition allows us to characterize the spectrum of the Laplacian, using perturbation theory, and hence the behavior of several dynamic processes. In particular, we are able to assess the diffusion time scales in any multiplex structure, and we can also infer the optimal value of the synchronization ratio in terms of the master stability function.

The paper is structured as follows: in Sec.~II, we present the structural decomposition of the Laplacian of the multiplex (from now on supra-Laplacian) into intralayer and interlayer networks contributions; in Sec.~III, we analyze the role of the interlayer network; Sec.~IV is devoted to the perturbative analysis of the eigenvectors of the supra-Laplacian; in Sec.~V, we expose the implications of the findings in terms of the physics of dynamical processes in multiplex networks, and finally we state the conclusions.

%%%%%%%%%%%%%%%%%%%%%%%%%%%%%%%%%%%%%%%%%%%%%%%%%%%%%%%%%%%%%%%%
\section{Multiplex supra-Laplacian matrix}

Let us consider a multiplex network consisting of $M$~layers and $N$~nodes per layer. The intralayer connectivity of layer~$\alpha$ is expressed as an adjacency or strengths matrix $W^{(\alpha)} \in \mathbb{R}^{N \times N}$ whose corresponding Laplacian is $\laplintra{\alpha}=S^{(\alpha)}-W^{(\alpha)}$, where $S^{(\alpha)}$ is the diagonal matrix of the nodes' intralayer strengths. In multiplex networks it is supposed that the interlayer connectivity is identical for all nodes \cite{gomez13}, thus we may define the interlayer network $W^I \in \mathbb{R}^{M \times M}$ whose components represent the strength of the connection between every pair of layers, and the associated interlayer Laplacian is $\laplinter=S^I-W^I$. For the sake of simplicity, we will assume that the interlayer and intralayer networks are undirected, globally connected and without self-loops.

The supra-Laplacian $\supralapl$ of the whole multiplex \cite{gomez13} may be separated in two contributions,
\begin{eqnarray}
  \supralapl = \laplmultintra + \laplmultinter,
  \label{eq:supraLaplacian}
\end{eqnarray}
where $\laplmultintra$ stands for the supra-Laplacian of the independent layers and $\laplmultinter$ for the interlayer supra-Laplacian. The first one is just the direct sum of the intralayer Laplacians,
\begin{eqnarray}
  \laplmultintra=\left(
  \begin{array}{cccc}
    \laplintra{1} & 0 & \dots & 0\\
    0 & \laplintra{2} & \dots & 0\\
    \vdots & \vdots & \ddots & \vdots\\
    0 & 0 & \dots & \laplintra{M}\\
  \end{array}
  \right) = \bigoplus_{\alpha=1}^M \laplintra{\alpha},
  \label{eq:intraLaplacian}
\end{eqnarray}
while the interlayer supra-Laplacian may be expressed as the Kronecker (or tensorial) product of the interlayer Laplacian and the $N \times N$ identity matrix $I$,
\begin{eqnarray}
  \laplmultinter = \laplinter\otimes I.
  \label{eq:interLaplacian}
\end{eqnarray}

The decomposition of the supra-Laplacian given in Eq.~(\ref{eq:supraLaplacian}) is fundamental for the discovery of several spectral properties of the multiplex, which will be uncovered in the following sections.

%%%%%%%%%%%%%%%%%%%%%%%%%%%%%%%%%%%%%%%%%%%%%%%%%%%%%%%%%%%%%%%%
\section{The role of the interlayer network \label{sec:linearEigenvalues}}

It is well-known that the spectrum of the Kronecker product of two matrices is formed by the products of the eigenvalues of the individual matrices, and the associated eigenvectors are obtained by the Kronecker products of the eigenvectors. We can apply this property to Eq.~(\ref{eq:interLaplacian}), thus the eigenvalues of $\laplmultinter$ are equal to the eigenvalues of the interlayer network Laplacian $\laplinter$, since the identity matrix $I$ only has eigenvalues equal to one. Moreover, the multiplicity of each eigenvalue of $\laplmultinter$ is just $N$ times its multiplicity in $\laplinter$.

Let $\mathbf{x}^{I} \in \mathbb{R}^{M}$ be any of the eigenvectors of $\laplinter$, with eigenvalue $\lambda$. The vector $\mathbf{x} \equiv \mathbf{x}^{I}\otimes\mathbf{1}$, being $\mathbf{1}\equiv(1,\ldots,1)^{T} \in \mathbb{R}^{N}$, is an eigenvector of $\laplmultinter$ with eigenvalue $\lambda$, as explained above, but it is also an eigenvector of $\laplmultintra$ with zero eigenvalue:
\begin{equation}
  \laplmultintra \mathbf{x}
  = \( \bigoplus_{\alpha=1}^M \laplintra{\alpha}\) (\mathbf{x}^{I} \otimes \mathbf{1})
  = \bigoplus_{\alpha=1}^M x_{\alpha}^I (\laplintra{\alpha} \mathbf{1})
  = \mathbf{0}.
\end{equation}
Therefore,
\begin{equation}
  \supralapl \mathbf{x} = (\laplmultintra + \laplmultinter)\mathbf{x} = \lambda \mathbf{x},
\end{equation}
i.e.\ $\mathbf{x}$ is an eigenvector of the full supra-Laplacian. Consequently, denoting by $\Lambda(\laplinter) = \{\lambda^I_1=0, \lambda^I_2,\ldots,\lambda^I_{M}\}$ the spectrum of the interlayer Laplacian, with the eigenvalues sorted in ascending order, $\lambda^I_1\leq \lambda^I_2 \leq \cdots \leq \lambda^I_{M}$, we have shown that,
\begin{equation}
  \Lambda(\laplinter) \subset \Lambda(\supralapl)\,.
  \label{eq:intraeigssubset}
\end{equation}
Namely, all eigenvalues of the interlayer Laplacian are also eigenvalues of the supra-Laplacian. In general, these eigenvalues cannot be easily calculated and numeric computation is needed. In the following subsections we analyze several particular cases in which they can be easily computed.

%%%%%%%%%%%%%%%
\subsection{Solvable case: small number of layers}

When the number of layers is small it is possible to derive analytical expressions for the eigenvalues of $\laplinter$. For example, if there are two layers, the interlayer network is characterized by,
\begin{equation}
  W^I = \(
  \begin{array}{cc}
    0   & D_x \\
    D_x & 0
  \end{array}
  \), \ \
  \laplinter = \(
  \begin{array}{rr}
     D_x & -D_x \\
    -D_x &  D_x
  \end{array}
  \),
\end{equation}
with spectrum $\Lambda(\laplinter) = \{0, 2 D_x\}$, where $D_x$
is a parameter that controls the relative strength between the interlayer and the intralayer contributions,
\begin{equation}
  W^I = \(
  \begin{array}{ccc}
    0 & D_{12} & D_{13} \\
    D_{12} & 0 & D_{23} \\
    D_{13} & D_{23} & 0 \\
  \end{array}
  \),
\end{equation}
and then the non-zero eigenvalues $\lambda^I_2$ and $\lambda^I_3$ are given by
\begin{equation}
  \sum_{\alpha<\beta} D_{\alpha\beta}
  \pm \sqrt{
    \sum_{\alpha<\beta} \frac{D_{\alpha\beta}^3 - D_{12} D_{13} D_{23}}{D_{\alpha\beta}}
  },
\end{equation}
where $D_{a\beta}$ ($a,\beta=1,2,3$) denotes the weight of the links between the layers $a$ and $\beta$.
With four and five layers this kind of analytical expressions also exist (e.g.\ based on Cardano's method to solve cubic and quartic equations), but they are too involved to be useful. And for six or more layers, no general algebraic expressions exist due to the limitations imposed by Galois theorem.

%%%%%%%%%%%%%%%
\subsection{Solvable case: Uniform interlayer weights}

Another solvable case is the one in which the interlayer network is fully connected (without self-loops) and with all the weights equal to the same value, $w_{\alpha\beta}^I = D_x$ for all $\alpha \neq \beta$. Now $\Lambda(\laplinter) = \{0,M D_x,\ldots,M D_x\}$, where the non-zero eigenvalue $M D_x$ has multiplicity $M-1$.

If the interlayer network is uniform but not fully connected, all the eigenvalues are proportional to the common weight $D_x$, but the coefficients depend on the precise chosen topology. Many of them are also solvable, e.g.\ regular graphs, cycles, paths, etc.

%In general, however, the only way we found to discover general features of the supra-Laplacian spectrum is by using perturbation theory. In the next section we show these characteristics, which turn out to be specially useful to understand physical phenomena on multiplex networks.

%%%%%%%%%%%%%%%%%%%%%%%%%%%%%%%%%%%%%%%%%%%%%%%%%%%%%%%%%%%%%%%%
\section{Weak and strong interlayer interactions: perturbative analysis \label{sec:perturbSection}}

In the previous section we have shown that the eigenvalues of the interlayer network are also eigenvalues of the full supra-Laplacian. Here we show, using perturbation theory, that these eigenvalues also impose important restrictions on the shape of the whole spectrum of the supra-Laplacian. These characteristics are specially useful to understand physical phenomena on multiplex networks.

To account for the relative strength between the inter- and intralayer connections, we denote by $D_x$ the quotient between their maximum values, thus we can write,
\begin{equation}
	\supralapl = \laplmultintra + D_{x} \hat{\laplmultinter} = \laplmultintra + D_{x} \hat{\laplinter}\otimes I,
	\label{eq:supralapdx}
\end{equation}
where
\begin{equation}
	\laplmultinter = D_x \hat{\laplmultinter},\ \ \
	\laplinter = D_x \hat{\laplinter},\ \ \
	W^I = D_x \hat{W^I},
\end{equation}
and the largest components of $\laplmultintra$ and $\hat{\laplinter}$ are of the same order of magnitude. From now on we will use hats above objects where a factor $D_x$ has been extracted.

According to Eqs.~(\ref{eq:intraeigssubset}) and~(\ref{eq:supralapdx}), the spectrum of the supra-Laplacian $\supralapl$ contains the eigenvalues
\begin{equation}
  \Lambda(\laplinter) = \{\hat{\lambda}^I_1=0, \hat{\lambda}^I_2 D_x,\ldots,\hat{\lambda}^I_{M} D_x\} \subset \Lambda(\supralapl),
  \label{eq:lineareigs}
\end{equation}
where $\hat{\lambda}^I_{\alpha}$ are the eigenvalues of $\hat{\laplinter}$. On the other hand, the eigenvalues of the intralayer supra-Laplacian $\laplmultintra$ in Eq.~(\ref{eq:intraLaplacian}) are the union of the eigenvalues of the intralayer networks,
\begin{equation}
  \Lambda(\laplmultintra) = \bigcup_{\alpha=1}^M \Lambda(\laplintra{\alpha}).
  \label{eq:intraeigs}
\end{equation}

%%%%%%%%%%%%%%%
\subsection{Weak interlayer networks}
\label{sec:weak}

For small values of the interlayer network, $D_x\ll 1$, the smallest eigenvalues are those given in Eq.~(\ref{eq:lineareigs}), while the largest ones are perturbations of the non-zero intralayer eigenvalues in Eq.~(\ref{eq:intraeigs}). For the calculation of the shape of these perturbations, let us select any of the eigenvectors $\mathbf{x}^{(\gamma)} \in \mathbb{R}^{N}$ of the $\gamma$-th~intralayer Laplacian,
\begin{equation}
  \laplintra{\gamma} \mathbf{x}^{(\gamma)} = \lambda^{(\gamma)} \mathbf{x}^{(\gamma)}.
\end{equation}
The corresponding eigenvector of the intralayer supra-Laplacian is,
$\mathbf{v}^{(\gamma)} = \mathbf{e}_{\gamma} \otimes \mathbf{x}^{(\gamma)}$,
\begin{equation}
  \laplmultintra \mathbf{v}^{(\gamma)} = \lambda^{(\gamma)} \mathbf{v}^{(\gamma)},
  \label{eq:vzerosmall}
\end{equation}
where $\mathbf{e}_{\gamma} \in \mathbb{R}^{M}$ denotes the canonical vector with a unity in the $\gamma$-th component and zeros elsewhere. By substituting the first order perturbation,
\begin{eqnarray}
  \mathbf{v} & \approx & \mathbf{v}^{(\gamma)} + D_x \mathbf{v}' \\
  \lambda & \approx & \lambda^{(\gamma)} + D_x \lambda'
\end{eqnarray}
and Eq.~(\ref{eq:supralapdx}) into the eigenvalue equation,
\begin{equation}
  \supralapl \mathbf{v} = \lambda \mathbf{v},
\end{equation}
we recover Eq.~(\ref{eq:vzerosmall}) at $0$-th order, and,
\begin{equation}
  \laplmultintra \mathbf{v}' + \hat{\laplmultinter} \mathbf{v}^{(\gamma)} =
  \lambda^{(\gamma)} \mathbf{v}' + \lambda' \mathbf{v}^{(\gamma)}\,,
\end{equation}
at $O(D_x)$. Multiplying both sides by the left with the transpose of $\mathbf{v}^{(\gamma)}$, and taking into account the symmetry of the supra-Laplacians and Eq.~(\ref{eq:vzerosmall}), we get
\begin{equation}
  \lambda'
  = \frac{\mathbf{v}^{(\gamma)T} \hat{\laplmultinter} \mathbf{v}^{(\gamma)}}{\mathbf{v}^{(\gamma)T} \mathbf{v}^{(\gamma)}}
  = \frac{\hat{\ell}^I_{\gamma\gamma} \mathbf{x}^{(\gamma)T} \mathbf{x}^{(\gamma)}}{\mathbf{x}^{(\gamma)T} \mathbf{x}^{(\gamma)}}
  = \hat{\ell}^I_{\gamma\gamma}
\end{equation}
where $\hat{\ell}^I_{\alpha\beta}$ denotes the $(\alpha,\beta)$-component of the interlayer Laplacian $\hat{\laplinter}$. Hence, the first order perturbation of the $\gamma$-th~intralayer Laplacian eigenvalues at $D_x\ll 1$ are given by
\begin{eqnarray}
  \lambda & \approx & \lambda^{(\gamma)} + \hat{s}^I_{\gamma} D_x \label{eq:smalldxlambdas} \\
  \hat{s}^I_{\gamma} & = & \ds \sum_{\beta} \hat{w}^I_{\gamma\beta} \label{eq:smalldxlambdasshift}
\end{eqnarray}
where $\hat{\ell}^I_{\gamma\gamma}=\hat{s}^I_{\gamma}$ represents the strength of node $\gamma$ in the interlayer network $\hat{W}^I$. This means that, for small values of $D_x$, all the non-zero eigenvalues of the intralayer networks are shifted by a value $\hat{s}^I_{\gamma} D_x$, which only depends on the layers. For the zero eigenvalues no perturbation analysis is needed since we know the exact values, given in Eq.~(\ref{eq:lineareigs}).

%%%%%%%%%%%%%%%
\subsection{Strong interlayer networks}
\label{sec:strong}

The analysis in the limit of strong interlayer interactions, $D_x \gg 1$, is analogous to the weak interactions case. Defining $\epsilon = 1/D_x$, we can write the supra-Laplacian as,
\begin{equation}
  \supralapl = D_x ( \hat{\laplmultinter} + \epsilon\laplmultintra ) = D_x \hat{\supralapl},
\end{equation}
and consider the behavior at $\epsilon \ll 1$. The eigenvectors and eigenvalues to be perturbed are those from the interlayer supra-Laplacian $\hat{\supralapl}$. Calling $\mathbf{x}^{I}$ and $\hat{\lambda}^{I}$ any eigenvector and eigenvalue pair of $\hat{\laplinter}$,
\begin{equation}
  \hat{\laplinter} \mathbf{x}^{I} = \hat{\lambda}^{I} \mathbf{x}^{I},
  \label{eq:hatlaplinter}
\end{equation}
and using Eq.~(\ref{eq:interLaplacian}), we realize that, for any vector $\mathbf{u} \in \mathbb{R}^{N}$,
\begin{equation}
  \hat{\laplmultinter} (\mathbf{x}^{I} \otimes \mathbf{u}) = (\hat{\laplinter} \otimes I) (\mathbf{x}^{I} \otimes \mathbf{u})
  = \hat{\lambda}^{I} (\mathbf{x}^{I} \otimes \mathbf{u}).
\end{equation}
Therefore, we apply the following perturbation,
\begin{eqnarray}
  \mathbf{v} & \approx & \mathbf{x}^{I} \otimes \mathbf{u} + \epsilon \mathbf{v}', \\
  \hat{\lambda} & \approx & \hat{\lambda}^{I} + \epsilon \hat{\lambda}',
\end{eqnarray}
which substituting in
\begin{equation}
  \hat{\supralapl} \mathbf{v} = \hat{\lambda} \mathbf{v},
\end{equation}
leads, at $O(\epsilon)$, to
\begin{equation}
  (\hat{\laplinter} \otimes I) \mathbf{v}' + \laplmultintra (\mathbf{x}^{I} \otimes \mathbf{u})
  = \hat{\lambda}^{I} \mathbf{v}' + \hat{\lambda}' (\mathbf{x}^{I} \otimes \mathbf{u}).
\end{equation}
Its $\alpha$~block row is equal to
\begin{equation}
  \sum_{\beta} \hat{\ell}^I_{\alpha\beta} \mathbf{v}'_{\beta}
  + x^I_{\alpha} \laplintra{\alpha} \mathbf{u}
  = \hat{\lambda}^{I} \mathbf{v}'_{\alpha} + \hat{\lambda}' x^I_{\alpha} \mathbf{u}.
\end{equation}
Now we multiply by $x^I_{\alpha}$, sum over $\alpha$
\begin{eqnarray}
  & &
  \sum_{\beta} \left( \sum_{\alpha} x^I_{\alpha} \hat{\ell}^I_{\alpha\beta} \right) \mathbf{v}'_{\beta}
  + \sum_{\alpha} (x^I_{\alpha})^2 \laplintra{\alpha} \mathbf{u} \nonumber \\
  & &
  = \hat{\lambda}^{I} \sum_{\alpha} x^I_{\alpha} \mathbf{v}'_{\alpha}
  + \hat{\lambda}' \left( \sum_{\alpha} (x^I_{\alpha})^2 \right) \mathbf{u},
\end{eqnarray}
and use Eq.~(\ref{eq:hatlaplinter}) to obtain
\begin{equation}
  \left[
    \frac{1}{\module{\mathbf{x}^{I}}^2} \sum_{\alpha} (x^I_{\alpha})^2 \laplintra{\alpha}
  \right] \mathbf{u} = \hat{\lambda}' \mathbf{u}.
  \label{eq:largedxperturb}
\end{equation}

Among the sorted eigenvalues
$\Lambda(\hat{\laplinter}) = \{\hat{\lambda}^I_1,\ldots,\hat{\lambda}^I_{M}\}$
it is convenient to study $\hat{\lambda}^I_1=0$ separately from the rest of the non-zero ones. In this case, we know that $\mathbf{x}^{I}=\mathbf{1}$, and Eq.~(\ref{eq:largedxperturb}) reduces to
\begin{equation}
  \left(
    \frac{1}{M} \sum_{\alpha} \laplintra{\alpha}
  \right) \mathbf{u} = \hat{\lambda}' \mathbf{u}.
  \label{eq:largedxaverage}
\end{equation}
This means that, for $D_x \gg 1$, the associated eigenvalues $\lambda$ of the supra-Laplacian $\supralapl$,
\begin{equation}
  \lambda \approx D_x ( 0 + \epsilon \hat{\lambda}' ) = \hat{\lambda}'
\end{equation}
are the eigenvalues of the Laplacian $L^{\mbox{\scriptsize AV}}$ of the {\em average network}
\begin{equation}
  W^{\mbox{\scriptsize AV}}=\frac{1}{M} \sum_{\alpha} W_{\alpha}.
  \label{eq:avgnet}
\end{equation}
This result was introduced in \cite{gomez13} only for the particular case of the two-layer multiplex, $M=2$.

For the non-zero eigenvalues $\hat{\lambda}^I \neq 0$
\begin{equation}
  \lambda \approx D_x (\hat{\lambda}^I  + \epsilon \hat{\lambda}' ) = \hat{\lambda}^I D_x + \hat{\lambda}'
  \label{eq:asymptoticeigs}
\end{equation}
therefore the corresponding eigenvalues of the supra-Laplacian diverge linearly with $D_x$, with shifts given by the eigenvalues of Eq.~(\ref{eq:largedxperturb}). One particular shift is $\hat{\lambda}'=0$ when $\mathbf{u} = \mathbf{1}$, thus recovering the previously found exact eigenvalues given in Eq.~(\ref{eq:lineareigs}).

%%%%%%%%%%%%%%%
\subsection{Global structure of the supra-Laplacian spectrum}
\label{sec:sumarySpectr}

The global picture of the supra-Laplacian spectrum, which can be deduced from the previous subsections, may be summarized in the following list:
\begin{itemize}
\item There is one exact eigenvalue $\lambda=0$ for all values of $D_x$.
\item There are $M-1$ non-zero exact eigenvalues, $\{\hat{\lambda}^I_2 D_x,\ldots,\hat{\lambda}^I_{M} D_x\}$, which are linear in $D_x$.
\item For $D_x\ll 1$, the smallest non-zero eigenvalues are the linear ones, which are $O(D_x)$. The rest $M(N-1)$ eigenvalues are $O(1)$, and are given by Eq.~(\ref{eq:smalldxlambdas}) and~(\ref{eq:smalldxlambdasshift}).
\item For $D_x\gg 1$, the $O(1)$ and smallest $N-1$ non-zero eigenvalues are approximately the non-zero eigenvalues of the Laplacian of the average network in Eq.~(\ref{eq:avgnet}). The rest $N(M-1)$ are $O(D_x)$ as given in Eq.~(\ref{eq:asymptoticeigs}), including the exact linear eigenvalues.
\item The eigenvalues are continuous and non-decreasing functions of $D_x$.
\end{itemize}
This structure is completely general provided the interlayer and intralayer networks are undirected, connected and without self-loops. The changes produced by breaking the last two conditions are not important, but for directed networks complex eigenvalues may appear, thus modifying significantly the properties of the supra-Laplacian spectrum.

\begin{figure}[!t]
  \begin{center}
  \mbox{\includegraphics[width=\columnwidth]{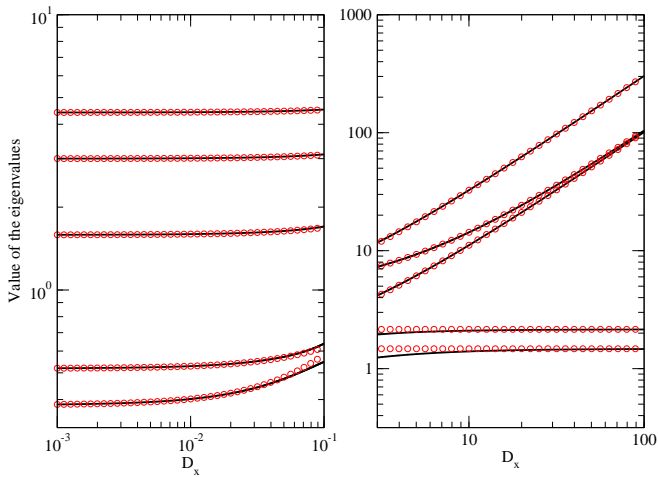}}
  \end{center}
  \caption{(color online) Plot of five eigenvalues of a multiplex of 3 layers, together with the proposed approximation, for weak and strong interlayer network. Each layer corresponds to a toy network of five nodes. Solid lines are the values of the eigenvalues and red circles the analytical approximation. Left panel: weak coupling. Right panel: strong coupling.}
  \label{fig:eigApprox}
\end{figure}

Figure~\ref{fig:eigApprox} illustrates the accuracy of the approximation for weak and strong interlayer networks. The figure shows a subset of the eigenvalues, together with the proposed approximation, of a multiplex network of tree layers. In each layer we have a different toy network of five nodes.

In the following section we will present the exploitation of these results in the context of physical processes running on top of general multiplex networks.

%%%%%%%%%%%%%%%%%%%%%%%%%%%%%%%%%%%%%%%%%%%%%%%%%%%%%%%%%%%%%%%%
\section{Physical implications of the spectrum of the Laplacian of multiplex networks}

%%%%%%%%%%%%%%%
\subsection{On the timescale of diffusion dynamics}
\label{sec:diffusionAnalysis}

Equivalently to \cite{gomez13}, we consider a particular diffusion dynamics where nodes are linearly coupled between them. Under the context of a multiplex network, as commented in the introduction, this coupling is differentiated between intralayer and interlayer. We consider that the interlayer coupling constant is the same for all nodes between two layers. Under this setting, it is possible to model the evolution of the states of node $x_{i\alpha}$ (node $i$ at layer $\alpha$) with the following differential equation:
\begin{eqnarray}
\label{eq:diff}
\dot{x}_{i\alpha} = \sum_{j=1}^{N}w_{ij}^{(\alpha)}(x_{j\alpha}-x_{i\alpha}) + \sum_{\beta=1}^{M}\ell_{\alpha\beta}(x_{i\beta}-x_{i\alpha}),
\end{eqnarray}
where $w_{i,j}^{(\alpha)}$ is the weight of the connectivity between nodes $x_{i\alpha}$ and $x_{j\alpha}$ and $\ell_{\alpha\beta}$ is the interlayer coupling constant.

The diffusion equation defined by Eq.~(\ref{eq:diff}) can be represented in matrix form by:
\begin{eqnarray}
\dot{\mathbf{x}}=-\(\laplmultintra + D_{x}\hat{\laplmultinter}\)\mathbf{x} =-\supralapl\mathbf{x}, \label{eq:diffusionMatForm}
\end{eqnarray}
where $\laplmultintra$ and $\laplmultinter=D_{x}\hat{\laplmultinter}$ represent the intralayer and interlayer supra-Laplacians and $D_{x}$ represents the quotient defined in Sec.~\ref{sec:perturbSection}.

The solution of this equation in terms of normal modes is given by $\phi_i(t)=\phi(0)e^{-\lambda_{i}t}$, where $\lambda_i$ are the eigenvalues of the supra-Laplacian matrix $\supralapl$. We can see that the eigenvalue that governs the convergence of the process is given by the second eigenvalue. Concluding that the time scale for the diffusion process becomes $\tau\propto\lambda_{2}^{-1}$.

For small couplings between the different layers (i.e. $D_{x}\ll 1$) $\lambda_{2}\(\supralapl\)$ will correspond to the second eigenvalue of the interlayer network, $\laplmultinter$. In that case, $\tau \propto (\hat{\lambda}_{2}^{I}D_{x})^{-1}$ (see Sec.~\ref{sec:weak}).

For large interlayer coupling (i.e. $D_{x}\gg 1$), $\lambda_{2}\(\supralapl\)$ can be approximated by the second eigenvalue of the average network defined in Eq.~(\ref{eq:avgnet}). Thus for large $D_{x}$ values $\tau \propto (\lambda_{2}(W^{\mbox{\scriptsize AV}}))^{-1}$ (see Sec.~\ref{sec:strong}).

\begin{figure}[!t]
  \begin{center}
  \mbox{\includegraphics[width=\columnwidth]{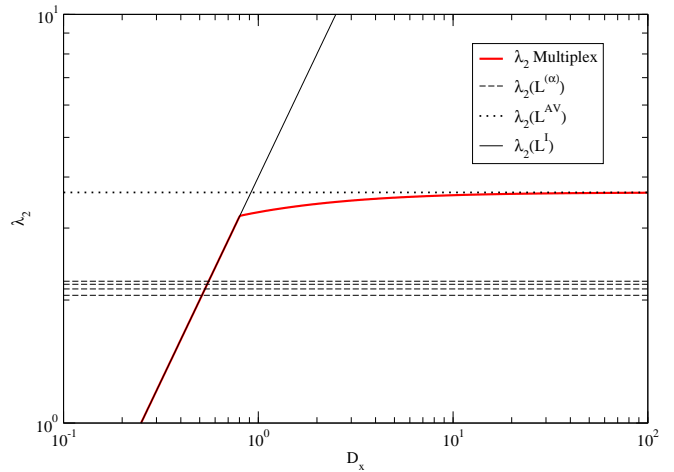}}
  \end{center}
  \caption{(color online) Comparison between the second eigenvalue of the different laplacians for a multiplex of four layers. Each layer contains a power-law degree distribution network of $200$ nodes generated using the Bar\'{a}basi-Albert model. Other network configurations can be found in the Supplemental material.}
  \label{fig:eigDiffu}
\end{figure}

Figure~\ref{fig:eigDiffu} shows the wellness of the proposed approximation for small and large $D_{x}$ coefficients for a multiplex of four layers. See Supplemental Material for more examples.

%%%%%%%%%%%%%%%
\subsection{On the stability of the synchronization manifold of coupled phase oscillators}
\label{sec:eigenratio}

Now, we focus on another ubiquitous physical process, the synchronization of phase oscillators in networks \cite{arenas08}.
Let us assume that the phase oscillators are embedded in the nodes of a multiplex network structure, and that the phases (states) are different
at different layers, meaning that we have a multi-phase oscillator. Extending previous results on the synchronizability (or strictly speaking, the stability of the synchronization manifold) to multiplex and taking advantage of the Master Stability Function \cite{pecora98,barahona02}, we reduce the problem of assessing synchronizability to that of computing the eigenratio $R=\lambda_{N}/\lambda_{2}$, in the multiplex network, where $\lambda_N$ is the largest eigenvalue. In particular, we analyze the asymptotic behavior of $R$ for the cases of weak interlayer coupling values ($D_x \ll 1$) and strong interlayer coupling values ($D_x \gg 1$).

For weak interlayer coupling values, as Sec.~\ref{sec:weak} states, the second eigenvalue of the supra-Laplacian becomes the second eigenvalue of the network of layers, i.e. $\lambda_{2}(\supralapl)\approx\lambda_{2}(\laplmultinter)$. The largest eigenvalue in this case corresponds to the largest eigenvalue of the Laplacians of the different layers, i.e. $\lambda_{\max}(\supralapl)=\max\limits_{\alpha}(\lambda_{\max}(\laplintra{\alpha})+\hat{s}^{I}_{\alpha}D_{x}) $. Thus, the eigenratio can be approximated as,
\begin{equation}
R\approx\frac{\max\limits_{\alpha}(\lambda_{\max}(\laplintra{\alpha})+\hat{s}^{I}_{\alpha}D_{x})}{\lambda_{2}(\hat{\laplmultinter})D_{x}}.
\label{eq:leftratio}
\end{equation}

\begin{figure}[!t]
  \begin{center}
  \mbox{\includegraphics[width=\columnwidth]{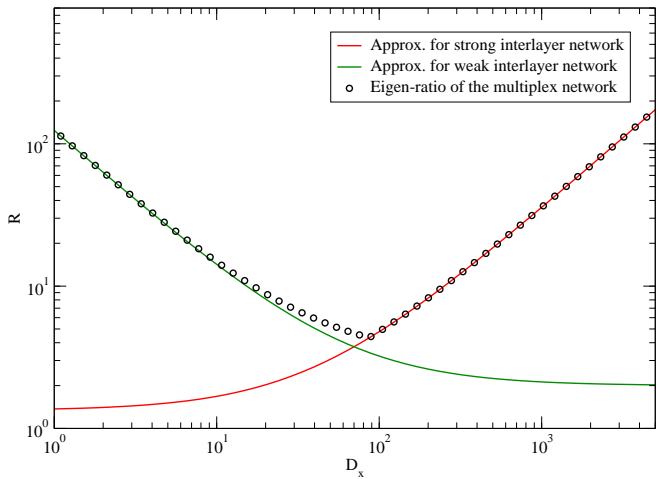}}
  \end{center}
  \caption{(color online) Plot of the eigenratio $R$ as a function of $D_x$. Circles corresponds to the real solution of the eigenratio, while solid lines correspond to the weak and strong analytical approximations. Note that the determination of the optimal value for the synchronizability of the system is extremely well approximated by the crossing of both analytical curves. Each layer layer corresponds to Erd\"{o}s-R\'{e}nyi networks of $200$ nodes and edge probability of $0.5$. Other network configurations can be found in the Supplemental material.}
  \label{fig:ratioER}
\end{figure}

For strong interlayer coupling the second eigenvalue can be approximated by the second eigenvalue of the average network, that is $L^{\mbox{\scriptsize AV}}$ (see Sec.~\ref{sec:strong}). The largest eigenvalue, in this case, is a function of the largest eigenvalue of the network of layers, $\lambda_{\max}(L^I)$, with an additional offset. This offset depends on the individual layers, $L^{(\alpha)}$, weighted by the entries of the eigenvector ${\mathbf{x}'}^{I}$ corresponding to the maximum eigenvalue of the network of layers. Thus, the largest eigenvalue is
\begin{equation}
  D_{x}\lambda_{\max}(\hat{L}^I) + \lambda_{\max}(L^{\mbox{\scriptsize WA}}_{\max}),
\end{equation}
where the laplacian $L^{\mbox{\scriptsize WA}}_{\max}$ is given by
\begin{equation}
  L^{\mbox{\scriptsize WA}}_{\max} = \frac{1}{\module{{\mathbf{x}'}^{I}}^2} \sum_{\alpha} ({x'}^{I}_{\alpha})^2 \laplintra{\alpha}.
\end{equation}
 Thus, the eigenratio for large interlayer coupling can be approximated as,
\begin{equation}
  R\approx\frac{D_{x}\lambda_{\max}(\hat{L}^I) + \lambda_{\max}(L^{\mbox{\scriptsize WA}}_{\max})}{\lambda_{2}(L^{\mbox{\scriptsize AV}})}.
\end{equation}
To illustrate the proposed approximation, we computed the eigenratio and the proposed approximation for a multiplex of four layers. Figure~\ref{fig:ratioER} shows the obtained results. See Supplemental material for additional plots with different multiplex topologies.

%%%%%%%%%%%%%%%%%%%%%%%%%%%%%%%%%%%%%%%%%%%%%%%%%%%%%%%%%%%%%%%%
\section{Conclusions}

We have presented the asymptotic analysis of the spectrum of the Laplacian of multiplex networks. We found analytical expressions that allow us to infer the behavior of dynamical processes, such as diffusion or synchronization, on top of multiplex networks. The findings reveal physical implications of the multiplex structure that have no counterpart in monoplex (one layer) networks. For example, in diffusive processes we find a super-diffusive behavior where the time scales of diffusion associated to the multiplex are shorter than in any particular individual layer network. In the analysis of the synchronizabity ratio $R$, we have found the existence of an optimal value of the coupling $D_x$ for which the synchronization of the full structure is the most stable. The applicability of the mathematical findings on the features of the Laplacian of multiplex networks sure go beyond this particular cases, and eventually can help to understand any process whose behavior is linked to the spectral properties of the Laplacian or other akin matrices.

\acknowledgments
This work has been partially supported by MINECO through Grant FIS2012-38266; by the EC FET-Proactive Project PLEXMATH (grant 317614), and the Generalitat de Catalunya 2009-SGR-838. A. A. also acknowledges partial financial support from the ICREA Academia and the James S.\ McDonnell Foundation. N. K and A. D. -G. acknowledges financial support from the EU/FP7-2012-STREP-318132 in the framework project LASAGNE.

%\bibliography{supra}

\appendix
\section{Supplemental material}

%\begin{figure}[!h]
  \begin{center}
  \mbox{\includegraphics[width=0.85\columnwidth]{./ratio3abN200.eps}}
  \end{center}
  Plot of the eigen-ratio and the proposed approximation for a multiplex of 3 layers. Each layers contains a scale-free network of $200$ nodes generated using the Bar\'{a}basi-Albert model.
%\end{figure}

%\begin{figure}[!h]
  \begin{center}
  \mbox{\includegraphics[width=0.85\columnwidth]{./ratio3co4P0_1N200.eps}}
  \end{center}
  Plot of the eigen-ratio and the proposed approximation for a multiplex of 3 layers. Each layers contains a network with 4 communities, each community corresponds to an Erd\"{o}s-R\'{e}nyi network with edge probability 0.5, and the inter-community edge probability is 0,1. The communities between different layers strongly overlap.
%\end{figure}
\vspace{2cm}

%\begin{figure}[!h]
  \begin{center}
  \mbox{\includegraphics[width=0.85\columnwidth]{./diffu4LayersER.eps}}
  \end{center}
  Comparison between the second eigenvalue of the different laplacians for a multiplex of 4 layers. Each layers contains an Erd\"{o}s-R\'{e}nyi of $200$ nodes with edge probability of 0.5.
%\end{figure}

\newpage

%\begin{figure}[!h]
  \begin{center}
  \mbox{\includegraphics[width=0.85\columnwidth]{./diffu4LayersCO4.eps}}
  \end{center}
  Comparison between the second eigenvalue of the different laplacians for a multiplex of 4 layers. Each layers contains a network with 4 communities, each community corresponds to an Erd\"{o}s-R\'{e}nyi network with edge probability 0.5, and the inter-community edge probability is 0,05. The communities between different layers strongly overlap.
%\end{figure}
\vspace{2cm}

%\begin{figure}[!h]
  \begin{center}
  \mbox{\includegraphics[width=0.85\columnwidth]{./diffu4Layers2CO_2AB.eps}}
  \end{center}
  Comparison between the second eigenvalue of the different laplacians for a multiplex of 4 layers. Two of the layers contain a network with 4 communities, each community corresponds to an Erd\"{o}s-R\'{e}nyi network with edge probability 0.5, and the inter-community edge probability is 0,05. The two communities strongly overlap. The other two layers contain an Erd\"{o}s-R\'{e}nyi network of $200$ nodes with edge probability of 0.5.
%\end{figure}

%\begin{figure}[!h]
  \begin{center}
  \mbox{\includegraphics[width=0.85\columnwidth]{./diffu4Layers3AB_1ER.eps}}
  \end{center}
  Comparison between the second eigenvalue of the different laplacians for a multiplex of 4 layers. Three of the layers contain a scale-free network of $200$ nodes generated using the Bar\'{a}basi-Albert model. The other layer contain an Erd\"{o}s-R\'{e}nyi network of $200$ nodes with edge probability of 0.5.
%\end{figure}
\vspace{2cm}

%\begin{figure}[!h]
  \begin{center}
  \mbox{\includegraphics[width=0.85\columnwidth]{./diffu4Layers2abm3_2abm7.eps}}
  \end{center}
  Comparison between the second eigenvalue of the different laplacians for a multiplex of 4 layers. Each layers contains a scale-free network of $200$ nodes generated using the Bar\'{a}basi-Albert model. The first two layers have been generated attaching, at each step, the new node to a 3 existing nodes, the third and fourth layers attaching the new node to a 7 existing nodes.
%\end{figure}

\end{document}